# On the analogy between the restricted primitive model and capacitor circuits: semi-empirical alternatives for over- and underscreening in the calculation of mean ionic activity coefficients

*Andrés González de Castilla,* Simon Müller, Irina Smirnova*

*corresponding author (andres.gdc@tuhh.de)

*Hamburg University of Technology, TUHH, Institute of Thermal Separation Processes, Eißendorfer Straße 38 (O) 21073 Hamburg, Germany.*

## Abstract

The analogy between the restricted primitive model and capacitor circuits, originally described decades ago for the Mean Spherical Approximation, is explored to demonstrate its transferability in linearized electrolyte theories. On this basis, we offer an explanation of why treating the salt diameter as a free adjustable parameter blurs differences between electrolyte theories in the calculation of mean ionic activity coefficients. Furthermore, a capacitor circuit analogy with an approximation of the Dressed Ion Theory is applied to develop a modified closest approach parameter "$b$" for the Pitzer-Debye-Hückel term. This modification is able to account for the qualitative effects of over- and underscreening in the calculation of mean ionic activity coefficients. This is achieved by defining "$b$" as a semi-empirical function that allows close resemblance with the multiple decay-length extension of the Debye-Hückel theory for high dielectric constant values and that lies close to recommended literature values of "$b$" for low dielectric constant values. Finally, as proof of principle, this modified semi-empirical Pitzer-Debye-Hückel term is implemented in the predictive COSMO-RS-ES model, an own reimplementation of the COSMO-RS theory developed for thermodynamic property calculations of electrolyte systems. It is shown that the modified semi-empirical Pitzer-Debye-Hückel term is an effective replacement for the recently published version of COSMO-RS-ES with explicit considerations for ion pairing. This reduces the modelling complexity by implicitly considering ion pairing and improves overall qualitative performance for the prediction of salt solubilities in mixed-solvent systems and even mean ionic activity coefficients in non-aqueous media.

## Introduction

Electrolyte theory remains to this day a challenging topic of particular interest for a large number of applications in the chemical industry. The presence of ions, either with localized or delocalized charges, large or small, monoatomic or polyatomic, adds a significant degree of complexity to the calculation of thermodynamic properties. Several questions[1–5] regarding the nature of highly concentrated electrolytes are a present matter of discussion in diverse working groups. A systematic review of some of these was recently published[1], yet simultaneously more intricate details of electrolyte theory and additional questions regarding non-Debye screening continued to develop.[6–8] Addressing all of these is a difficult task that will require the cumulative work of many researchers around the world. Specific issues have been recently addressed like, for instance, the relevance of the Born term as discussed by Simonin[9] and the fundamentals of underscreening in the Debye-Hückel context as presented by Kjellander.[8,10]

As stated by Kontogeorgis et al.[1] *there are far too many unclear questions and concepts in electrolyte thermodynamics.* The most commonly applied expressions therefore remain the simpler models: the Debye-Hückel (DH) with a distance of closest approach,[11] the primitive Mean Spherical Approximation (MSA)[12,13] or the Pitzer-Debye-Hückel (PDH)[14,15] term. These expressions have been combined with equations of state (EOS) for electrolytes[16–23] as well as with liquid lattice models like the e-NRTL,[24,25] UNIFAC/UNIQUAC based electrolyte models[26–32] and COSMO-RS based electrolyte models.[33–40]

An interesting finding from recent literature is that the practical results obtained from the unrestricted primitive MSA and DH theories have apparently negligible differences when the salt diameter in DH is set to 5/6 of the MSA diameter.[1,3] In the present work, we explore the workings behind the aforementioned scaling of the salt diameter through the spherical capacitor analogy proposed by Blum[41,42] for linearized primitive electrolyte models. Circuit representations are provided for the DH theory, the restricted primitive MSA, the PDH term and the recently published[43] multiple decay-length extension of the DH theory (MDE-DH). Furthermore, by means of an analogy with a hypothetical capacitor circuit, a correction for underscreening in 1:1 electrolytes with very low dielectric constant values is introduced to modify the closest approach parameter of the PDH term. This development allows a reinterpretation of the recommended literature values[44] for the closest approach parameter. Finally, this modified PDH term is applied systematically with the predictive COSMO-RS-ES electrolyte model,[35] proving to be a considerably simpler and more consistent replacement of our recently published[37] iterative ion pairing algorithm.

## Theory and Capacitor Circuit Representations

To represent linearized electrolyte theories as capacitor circuits, we first require the excess internal energy from a coulombic potential of charges embedded in a spherical volume of some dielectric. This is expressed in terms of the Green formula, as suggested many years ago by Blum:[42]

$$\Delta E = \frac{Q^2}{2C} = \frac{2\pi e^2}{\varepsilon_0} \sum_{ij} \rho_i \rho_j z_i z_j \int_0^\infty \frac{r^2 g_{ij}(r)}{r} dr \qquad (1)$$

where $Q$ is the total charge of the spherical capacitor given by ions of type $i$ and counter-ions of type $j$, $\rho$ stands for number density, $z$ stands for ion valence and $g_{ij}(r)$ is the symmetric pair correlation function. Based on the previous equation one can express the energy density of ions of type $i$ as follows:[42]

$$\Delta E = \frac{Q^2}{2C} = -\frac{\sum_i \rho_i z_i^2 e^2}{2\epsilon} \cdot \left(\frac{1}{C^X}\right) \tag{2}$$

where for a medium with a dielectric constant $\varepsilon_r$, Gaussian units ($\epsilon = 4\pi\varepsilon_0\varepsilon_r$) will be commonly applied to follow the same convention as Blum[42] and Wei and Blum.[41] The capacitance is $C = \epsilon C^X$, where $C^X$ is the effective capacitance given by some electrolyte theory $X$. Equation (2) can be interpreted as a series and/or parallel circuits of capacitors and this is apparently valid for any linearized electrolyte theory. The simplest case, as shown by Blum,[42] is the DH limiting law (DHLL) obtained by introducing the pair correlation function from DH theory in equation (1) from which the contribution of ions of type $i$ to $\Delta E$ is:

$$\Delta E_i = -\frac{\rho_i z_i^2 e^2}{2\epsilon} \cdot \left(\frac{1}{1/\kappa_D}\right) = -\frac{\rho_i z_i^2 e^2 \kappa_D}{8\pi\varepsilon_0\varepsilon_r} \tag{3}$$

The effective capacitance would then be $C^{DHLL} = 1/\kappa_D$. The physical situation described by the DHLL is that of a point charge $i$ surrounded by an ion cloud of non-central point charges $j$ where significant $ij$ interactions take place within a single spherical capacitor of radius $r = \lambda_D = 1/\kappa_D$, as shown in Figure 1:

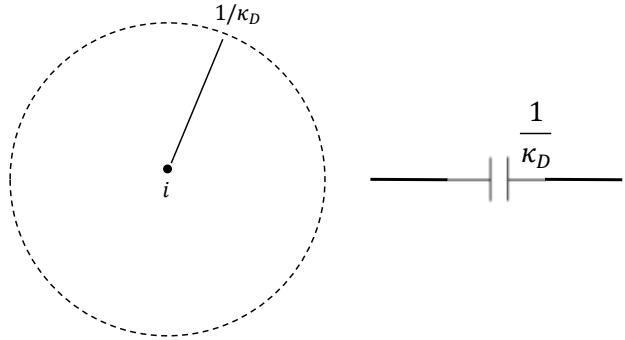

Figure 1. The physical situation in DHLL as described by a single capacitor.

### Debye-Hückel theory and the Mean Spherical Approximation

One of the major improvements starting from the DHLL was to consider a central sphere with a distance of closest approach $a$ (taken as the salt diameter).[11] Thereby the potential is divided in two regions: one where the interaction potential in a hard core region at $r < a$ and a coulombic potential with a Yukawa type exponential decay at $r \geq a$. As shown by Blum,[42] by applying equation (2), this arrangement of two concentric spheres can be represented for the DH and MSA theories as follows:

$$\Delta E_i = -\frac{\rho_i z_i^2 e^2}{2\epsilon} \cdot \left(\frac{1}{C^{DH}}\right) = -\frac{\rho_i z_i^2 e^2}{2\epsilon} \cdot \left(a + \frac{1}{\kappa_D}\right)^{-1} = -\frac{\rho_i z_i^2 e^2}{2\epsilon} \cdot \left(\frac{\kappa_D}{1 + \kappa_D a}\right) \quad (4)$$

$$\Delta E_i = -\frac{\rho_i z_i^2 e^2}{2\epsilon} \cdot \left(\frac{1}{C^{MSA}}\right) = -\frac{\rho_i z_i^2 e^2}{2\epsilon} \cdot \left(\frac{a}{2} + \frac{1}{2\Gamma}\right)^{-1} = -\frac{\rho_i z_i^2 e^2}{2\epsilon} \cdot \frac{2\Gamma}{1 + \Gamma \cdot a} \quad (5)$$

with the known definitions for the Debye and MSA screening parameters:

$$\kappa_D = \sqrt{\frac{4\pi N_A e^2 \sum_i c_i z_i^2}{\epsilon k_B T}} = \sqrt{\frac{8\pi N_A e^2 I^{(c)}}{\epsilon k_B T}} \quad (6)$$

$$2\Gamma = \frac{1}{a}\left(\sqrt{1 + 2\kappa_D a} - 1\right) \quad (7)$$

It is relevant to clarify that Blum[42] applied a different DH version where the spherical capacitor is given by $a/2$ instead of $a$. The use of $a$ is more common to account for volumetric correlations between the central ion $i$ and the non-central ions $j$, but DH does not account for volumetric correlations between non-central ions. The physical situations and circuit representations of equations (4) and (5) are shown in Figure 2:

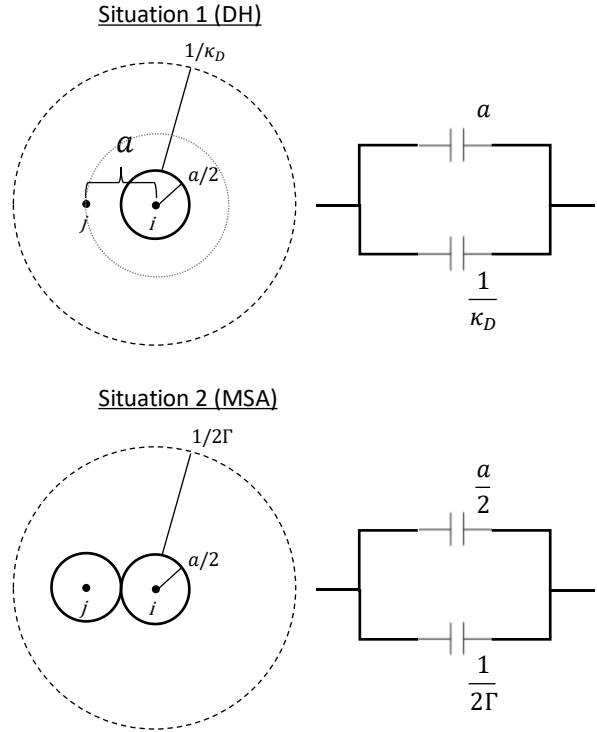

*Figure 2. The physical situation in DH theory and in MSA theory represented as two parallel effective capacitors.*

Volume exclusion effects in the DH context can be addressed by the Dressed Ion Theory (DIT)[45,46] and applied to conventional electrolytes with a recently published approximation of DIT: the MDE-DH theory.[43]

*Multiple Decay-Length Extension of the Debye-Hückel (MDE-DH) theory*

MDE-DH introduces the use of two decay lengths in DH theory: as electrolyte concentration rises, the decay length deviates from the theoretical Debye length and the electrostatic potential becomes oscillatory, indicating the presence of multiple decay lengths. This was first described by Kirkwood in the 1930s.[47] For deeper details the reader is referred to the original publication.[43]

In the DH derivation, the potential of the central ion as function of the radial distance $\psi_i(r)$ is given by solving the linearized Poisson-Boltzmann equation using the electro-neutrality condition to yield:

$$\psi_i(r) = \frac{q_i}{\epsilon} \cdot \frac{\exp(-\kappa_D r)}{r} \cdot \frac{\exp(\kappa_D a)}{1 + \kappa_D a} \qquad (8)$$

The dimensionless term $\exp(\kappa_D a)/(1+\kappa_D a)$ multiplied by an electron charge yields the effective charge $q_i^{eff}$ of the central ion. That is, the charge of the central ion and its ion cloud as seen by another charge at a given distance.[45] A renormalization of the screening length based on $q_i^{eff}$ yields the following relation:[8]

$$\frac{\kappa^2}{\kappa_D^2} = \frac{\frac{4\pi N_A \sum_i c_i q_i q_i^{eff}}{\epsilon k_B T}}{\frac{4\pi N_A \sum_i c_i q_i^2}{\epsilon k_B T}} = \frac{\exp(\kappa a)}{1 + \kappa a} \qquad (9)$$

Equation (9) must be solved numerically and yields two solutions, hereafter referred to as the two effective screening lengths $\kappa$ and $\kappa'$. These terms become complex valued above the Kirkwood crossover, where one becomes the complex conjugate of the other with:[8] $\kappa = \kappa_\Re + i\kappa_\Im$ and $\kappa' = \kappa_\Re - i\kappa_\Im$. The Kirkwood crossover is the point where the pair correlations transition into an oscillatory regime.

Two decay lengths imply multiple effective charges (e.g. $q$ and $q'$) and multiple effective dielectric screening parameters (e.g. $\mathcal{E}$ and $\mathcal{E}'$) to describe the dielectric response of the electrolyte. Consequently, an effective dielectric permittivity is given by $E_r^{eff} = \varepsilon_r \mathcal{E}$, where $\mathcal{E} = 1$ in the traditional DH formulation. Thus, for two decay lengths, the magnitude of the screened Coulomb potential divides into two effective contributions $E_r^{eff}$ and $E'^{eff}_r$.[8,43] The fundamental origin of these terms is described in the literature.[8,10,43]

In this work, we suggest the following circuit configuration for the MDE-DH theory, shown in Figure 3:

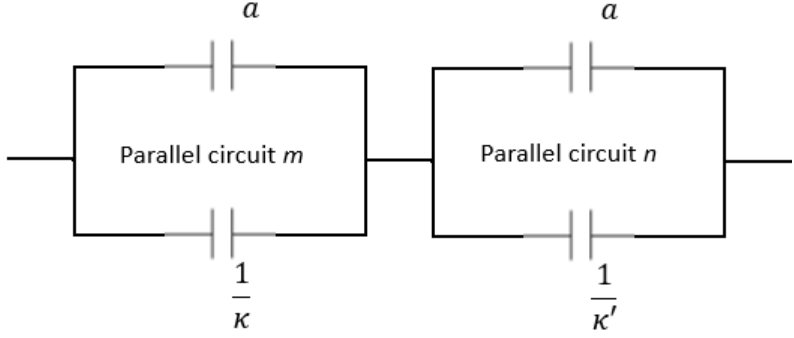

*Figure 3. MDE-DH shown as a series of two sets of parallel capacitors (m and n).*

MDE-DH therefore describes situation 1 in Figure 2, but with different decay lengths. In relation to capacitor circuit rules, the current (charges) and the voltage (energy density) are divided, and each contribution to the energy density has in turn two contributions from the central and cloud ions. Therefore, the configuration shown in Figure 3 is obtained. That is, a circuit where the voltage is divided (in series) and the current is divided (in parallel). By applying the rules of capacitor circuits one obtains the following effective capacitance:

$$\frac{1}{C^{MDE-DH}} = \frac{1}{\mathcal{E}\sum_m(C_m^{MDE-DH})} + \frac{1}{\mathcal{E}'\sum_n(C_n^{MDE-DH})} = \left(a\mathcal{E} + \frac{\mathcal{E}}{\kappa}\right)^{-1} + \left(a\mathcal{E}' + \frac{\mathcal{E}'}{\kappa'}\right)^{-1} \quad (10)$$

where the subscripts $m$ and $n$ refer to the corresponding parallel circuits in Figure 3. By introducing equation (10) in equation (2) one obtains:

$$\Delta E_i = -\frac{\rho_i z_i^2 e^2}{2\epsilon} \cdot \left(\frac{1}{\mathcal{E}} \cdot \left(a + \frac{1}{\kappa}\right)^{-1} + \frac{1}{\mathcal{E}'} \cdot \left(a + \frac{1}{\kappa'}\right)^{-1}\right) \quad (11)$$

which exactly corresponds to the MDE-DH equation below the Kirkwood crossover.[43] Above the Kirkwood crossover the previous equation reduces by means of the Euler rule and some algebra to:

$$\Delta E_i = -\frac{\rho_i z_i^2 e^2}{2\epsilon} \cdot \left[\frac{\kappa_\Re + \kappa_\Re^2 a + a\kappa_\Im^2 - \kappa_\Im \tan\vartheta}{(1 + \kappa_\Re a)^2 + a^2\kappa_\Im^2}\right] \quad (12)$$

which corresponds to the MDE-DH equation for $\Delta E_i$ above the Kirkwood cross-over and where the phase shift $\vartheta$ is a function of $\kappa_\Re$, $\kappa_\Im$ and $a$. The reader is referred to the original publication for further details.[43]

*Pitzer-Debye-Hückel*

The PDH model applies the LPB radial distribution function $g_{ij}(r)$ to the pressure equation, which in statistical mechanics expresses the osmotic pressure of a solution. Pitzer applied the same assumptions as the DH theory.[14,48] The first two terms of the Taylor expansion of $g_{ij}(r)$ are applied to the pressure equation. By applying the electroneutrality condition and integration by parts, the pressure equation reduces to the following result for the electrostatic contribution to the osmotic coefficient:[14,48]

$$\frac{\Pi^{el,1}}{\rho k_B T} - 1 = \Phi^{el} - 1 = -\frac{1}{24\pi\rho}\left(\frac{\kappa_D^3}{1+\kappa_D a}\right) \tag{13}$$

which can be introduced in the integral of the Gibbs-Duhem expression to yield an equation for the logarithmic activity coefficient with the following analytical solution:

$$\frac{\Delta E_i}{\rho_i} = -\frac{e^2 z_i^2}{2\epsilon}\left(\frac{1}{3}\right)\left\{\left(\frac{\kappa_D}{1+\kappa_D a}\right) + \frac{2}{a}\ln(\kappa_D a + 1)\right\} \tag{14}$$

For the chemical potential of the central ion with equation (14) the definition $\Delta A_i^E = \Delta E_i$, the assumption of $\Delta A_i^E \equiv \Delta G_i^E$ and the substitution with a closest approach parameter $b = a \cdot (\kappa_D/\sqrt{I^{(c)}})$ yields the commonly applied PDH term:[14,48]

$$\ln(\gamma_i^{el}) = -\left(\frac{z_i^2}{3}\right)(2\pi N_A d_s)^{\frac{1}{2}}\left(\frac{e^2}{4\pi\varepsilon\varepsilon_r k_B T}\right)^{\frac{3}{2}}\left\{\frac{\sqrt{I^{(m)}}}{1+b\sqrt{I^{(m)}}} + \frac{2}{b}\ln(1+b\sqrt{I^{(m)}})\right\} \tag{15}$$

where $I^{(m)}$ stands for ionic strength in the molality scale and $d_s$ is taken as the salt-free medium (solvent) density based on the assumption $I^{(c)} = I^{(m)}$.

Based on equation (2) and following the same rules for circuit capacitors, the corresponding capacitance for equation (14) is given by:

$$\frac{1}{C^{PDH}} = \left(3a + \frac{3}{\kappa_D}\right)^{-1} + \left(\frac{3a}{2\ln(\kappa_D a + 1)}\right)^{-1} \tag{16}$$

and its capacitor circuit configuration is:

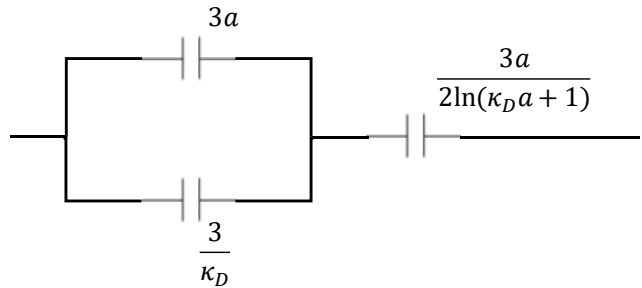

*Figure 4. DH from pressure described as a capacitor circuit.*

***Definition of over- and underscreening in electrolyte solutions***

Throughout this work, we will refer to the negative deviation of the Debye length from its theoretical Debye value arising from ion-ion correlations as *over-screening*. This phenomenon is explained by an increase of the leading screening parameter $\kappa$ that arises from ion-ion correlations (e.g. volume exclusion effects) and results in an increased effective charge of the ions.[4,8]

We will refer to the anomalous positive deviation of the Debye length from its theoretical value arising from ion-ion correlations as *underscreening*. This phenomenon is given by an anomalous decrease of the leading screening parameter $\kappa$ that arises from ion-ion correlations (e.g. dominant cation-anion coupling at high valences or very low dielectric constants) and results in a decreased effective ionic charges.[5,6,43,49] The present work addresses underscreening in low dielectric constants for 1:1 electrolytes. Deviations from high valences in symmetrical electrolytes are not considered. It is of relevance to clarify that only one type of underscreening is addressed in the present work. Underscreening may have multiple origins.[43]

## Methods

The present work is divided in two sections. Firstly, several linearized electrolyte models are compared on a theoretical basis and an explanation is offered of why scaling the salt diameter blurs differences between them. The principle behind this explanation is applied to develop a modified PDH model with semi-empirical corrections to qualitatively account for over- and underscreening effects in 1:1 electrolytes by means of a single universal parameter. In the second section, this modified PDH term is systematically applied in combination with the COSMO-RS-ES model. As commonly done in chemical engineering electrolyte $g^E$-models models, all calculations performed in both sections assume molarity equals molality.

*Comparison between electrolyte models: equations*

To compare results on a strictly theoretical level, the primitive II + IW model from Valiskó and Boda[50] is taken as the reference calculation using aqueous NaCl as an example. The II + IW model consists of primitive model Monte-Carlo calculations (II term) with solvation effects estimated from the Born treatment (IW term). Both terms have the salt concentration dependent dielectric constant of the solution as an input. Primitive model calculations are compared with the II + IW model given that Monte-Carlo methods are a good mathematical reference for the primitive model.

All linearized primitive model calculations of logarithmic activity coefficients are performed by means of the following expression:

$$\ln(\gamma_\pm) = \ln(\gamma_\pm^{el,X}) + \ln(\gamma_\pm^{Born}) + \ln(\gamma_\pm^{UHS}) \qquad (17)$$

where $\ln(\gamma_\pm^{el,X})$ denotes the electrostatic contribution obtained from electrolyte theory $X$ (DH, MSA, MDE-DH or the PDH term), $\ln(\gamma_\pm^{Born})$ is the contribution given by a Born term to describe the energy transfer between solvation in dilute and concentrated solutions and $\ln(\gamma_\pm^{UHS})$ represents the unrestricted hard-sphere term.

The models are compared for NaCl in water with $\varepsilon_r(c=0) = 78.35$ at 25°C. For the electrostatic terms a salt diameter of $a = 2.79$ Å (NaCl based on values from Marcus[51] with $r_{Na^+} = 0.98$ Å and $r_{Cl^-} = 1.81$ Å) is applied and the concentration dependent dielectric constant $\varepsilon_s(c)$ is modeled by means of the following experimental correlation for aqueous NaCl solutions from Buchner et al.:[52]

$$\varepsilon_s(c) = \varepsilon_r - 15.45c + 3.76c^{1.5} \tag{18}$$

where $c$ stands for molar concentration. The following Born term was applied:[50,53]

$$\ln(\gamma_i^{Born}) = -\frac{z_i^2 e^2}{8\pi R_i^{Born}\varepsilon_0 \kappa_B T}\left(\frac{1}{\varepsilon_r} - \frac{1}{\varepsilon_s(c)}\right) \tag{19}$$

For the Born term, the same ionic radii from Valiskó and Boda[50] from the original Grand-Canonical Monte-Carlo based II+IW simulations were taken: for the sodium cation a value of $R_{Na^+}^{Born} = 1.62$ Å and for the chloride anion a value of $R_{Cl^-}^{Born} = 2.26$ Å.

This result for the cation and anion is combined for a contribution to the mean ionic activity coefficient with the common expression $\gamma_\pm^{Born} = (\gamma_{cat}^{v_{cat}}\gamma_{an}^{v_{an}})^{\frac{1}{v}}$ where $v_i$ stands for the stoichiometric coefficient of ion $i$ and $v$ is the sum of the stoichiometric coefficients of the anion and the cation.

The hard-sphere term is given by an unrestricted hard sphere (UHS) contribution, for which the aforementioned radii from Marcus[51] are applied, and is given by the following set of equations:[54,55]

$$\gamma_\pm^{UHS} = (\gamma_+^{UHS}\gamma_-^{UHS})^{\frac{1}{v}} \tag{20}$$

$$\ln(\gamma_i^{UHS}) = \left(\delta_i - 1 + \frac{2\eta_i}{\eta}y_3\right)\ln(\Delta_{HS})$$
$$+ \frac{\eta}{\Delta_{HS}^2}\left[3(1-\alpha_i) + \delta_i + \frac{3\eta}{2}(\alpha_i - \beta_i - \delta_i - 1)\right] \tag{21}$$
$$+ \frac{\eta_i}{\Delta_{HS}^3}\left\{\eta\left[5y_3 - \frac{9}{2}y_1 - 2 + \eta\left(\frac{3}{2}y_1 - 3y_2 - 4y_3 + 1\right)\right] - 2y_3 + 4\right\}$$

with

$$\eta_i = \frac{\pi}{6}\rho\sigma_i^3 \quad \eta = \sum_i k_i\eta_i \quad k_i = \frac{\rho_i}{\rho} \quad \rho = \sum_i \rho_i$$

where $\rho_i$ is the number density of an ion with diameter $\sigma_i = 2r_i$ and with the following definitions:

$$\Delta_{HS} = 1 - \eta$$

$$y_1 = \sum_{i>j} \frac{\Delta_{ij}(\sigma_i + \sigma_j)}{(\sigma_i\sigma_j)^{1/2}}$$

$$y_2 = \sum_{i>j} \Delta_{ij} \sum_k \frac{k_k\eta_k}{\eta}\left(\frac{(\sigma_i\sigma_j)^{1/2}}{\sigma_k}\right)$$

$$y_3 = \left[\sum_i k_i\left(\frac{\eta_i}{\eta}\right)^{2/3}\right]^3$$

$$\Delta_{ij} = \frac{k_ik_j(\eta_i\eta_j)^{1/2}}{\eta}\left(\frac{(\sigma_i-\sigma_j)^2}{\sigma_i\sigma_j}\right)$$

$$\alpha_i = \frac{1}{k_i}\sum_k \frac{\sigma_i+\sigma_k}{(\sigma_i\sigma_k)^{1/2}}\Delta_{ik}$$

$$\beta_i = \sum_{i>j}\Delta_{ij}\frac{\eta_i}{\eta}\left(\frac{(\sigma_i\sigma_j)^{1/2}}{\sigma_i}\right) + \frac{1}{k_i}\sum_j \Delta_{ij}\sum_k k_k\frac{\eta_k}{\eta}\left(\frac{(\sigma_i\sigma_j)^{1/2}}{\sigma_k}\right)$$

$$\delta_i = 3\left(\frac{\eta_i}{\eta}y_3\right)^{2/3}$$

The selection of an unrestricted hard sphere term instead of a restricted hard sphere term like, for instance, the Carnahan-Starling term,[56] lies on the fact that size asymmetry effects are strongly hard-core dominated at low to moderate concentrations. Thus, while the theoretical inconsistency of combining a restricted electrostatic term with an unrestricted hard-core term is evident and should not be used for the pair correlations, in practical thermodynamic terms (calculation of activity coefficients alone) this inconsistency is perfectly negligible for small spherical ions at sufficiently low concentrations (e.g. $c < 5\ M$). A brief qualitative proof of this argument by means of comparisons with Hyper-Netted Chain (HNC) theory calculations from the literature is included in the Supporting Information. Findings from this section are used to present a modified closest approach parameter for the PDH term.

*Application with the COSMO-RS-ES model: computational details*

The COSMO-RS-ES model[35–37] combines the predictive $g^E$-model COSMO-RS[57] with universal equations for ion-ion and ion-solvent interactions including a PDH term for long ranged electrostatic interactions as shown in equation (22). The COSMO-RS based equations applied are those of our previous publication.[37] For a more detailed description of the model, the reader is referred to the COSMO-RS-ES literature.[35–37]

$$\ln(\gamma_i) = \ln(\gamma_i^{SR,COSMO-RS}) + \ln(\gamma_i^{el,PDH}) \qquad (22)$$

In our previous work[37] the PDH term is applied with an effective ionic strength that is iteratively calculated to qualitatively correct for underscreening (i.e. extensive ion pairing) in solutions with very low dielectric constants. This is achieved based on an approximation to the Bjerrum association constant.[37] In the present work the expression from Ebeling[58,59] was used. This theoretically more consistent expression in both the MSA and the DH context,[60] is given by:

$$K_{assoc}(T, a, \varepsilon_r) = 8\pi N_A a^3 \sum_{n=1}^{\infty} \frac{B^{2n+2}}{(2n+2)! \cdot (2n-1)} \qquad (23)$$

where we express volume in cubic meters (thus, $a$ is given in 10$^{-10}$ m) and with $B = \frac{\lambda_B}{a} = e^2/(\epsilon k_B T a)$ for a 1:1 electrolyte, where $\lambda_B$ is the Bjerrum length. $B$ is commonly referred to as the fluid specification. The salt diameter in the long-range term and in the calculation of the association constant is taken as a scaled value of the sum of the COSMO radii of the cation and anion.

The parameterization procedure, as well as the applied electrolyte database, remain the same as in our previous publication.[37] The parameterization is performed in two steps. Firstly, aqueous mean ionic activity coefficients (MIACs) of diverse alkali salts (1190 data points), liquid-liquid equilibria (LLE) of alkali salts in aqueous mixed solvent media (1126 data points) and Gibbs free energies of transfer of ions from water to organic solvents (992 points) are employed. In a second step, the parameterization is repeated with aqueous MIAC and LLE data only using the resulting values from the first step as starting values.

The COSMO-RS-ES parameters are optimized with a Levenberg-Marquardt algorithm to minimize the objective functions that have already been reported in our previous work.[35,36] The absolute average deviation (AAD) for the number of data points ($N_{DP}$) per system type is evaluated as follows:

$$AAD_{MIAC} = \frac{1}{N_{DP}} \sum_i \left| \ln\left(\gamma_{\pm,i}^{(m),*,exp}\right) - \ln\left(\gamma_{\pm,i}^{(m),*,calc}\right) \right| \qquad (24)$$

$$AAD_{LLE} = \frac{1}{N_{DP}} \sum_i \left| \ln\left(K_{salt,i}^{OS,exp}\right) - \ln\left(K_{salt,i}^{OS,calc}\right) \right| \qquad (25)$$

$$AAD_{G-transfer} = \frac{1}{N_{DP}} \sum_i \left| \frac{\Delta G_{t,i\ w \to s}^0}{RT} - \ln\left(\frac{\gamma_{i,s}^\infty}{\gamma_{i,w}^\infty}\right) \right| \qquad (26)$$

where $K_{salt}^{OS}$ is the partition coefficient of a salt between an salt-rich phase $S$ and an organic phase $O$; the subscript $(m)$ denotes molality and $\Delta G_{t,i\ w\rightarrow s}^0$ denotes a literature value for the energy of transfer of an ion between water and some organic solvent $s$. The results from aqueous MIAC, LLE and Gibbs free energies of transfer will be referred to as correlations. Data that does not belong to the training set will be referred to as predictions. For these SLE predictions, 835 data points for salt solubility in mixed and non-aqueous solvents is applied and the AAD is evaluated as follows:

$$AAD_{SLE} = \frac{1}{N_{DP}} \sum_i \left| ln\left(\gamma_{\pm,i,other}^{+,expected}\right) - ln\left(\gamma_{\pm,i,other}^{+,calc}\right) \right| \quad (27)$$

with

$$ln\left(\gamma_{\pm,i,other}^{+,expected}\right) = \ln\left(x_{\pm,ref} \cdot \gamma_{\pm,ref}^+\right) - \ln\left(x_{\pm,experiment}\right) \quad (28)$$

where the subscript $ref$ indicates a reference solvent which for our case is always water. As a measure of qualitative performance, the average relative deviation (ARD) for the SLE systems is also applied:

$$ARD_{SLE} = \frac{1}{N_{DP}} \sum_i \left| 1 - \frac{ln\left(\gamma_{\pm,i,other}^{+,calc}\right)}{ln\left(\gamma_{\pm,i,other}^{+,expected}\right)} \right| \quad (29)$$

Additional predictions of MIAC values in pure organic solvents are also presented for comparison. These were taken from the literature as osmotic coefficients and converted to activity coefficient values via the Gibbs-Duhem relation and/or the Pitzer equations when parameters were available. A list of the applied systems is included in the Supporting Information.

## Results

### *Comparison between electrolyte models*

Maribo-Mogensen et al.[3] reported that by scaling the salt diameter in DH ($a_{DH}$) by a factor of 5/6 with respect to that of the unrestricted primitive MSA theory ($a_{MSA}$) leads to a very similar performance. This would suggest that the choice to calculate activity coefficients should then most likely be dominated by the simplicity of the implementation when the salt diameter is taken as a freely adjustable parameter.

As a complement to their findings, one can first take a closer look at the simpler relations: equations (4) and (5). If one attempts to force an equality between the models (although these equations do not describe the same physical situation, as seen in Figure 2), then the following relation between the effective capacitances would have to be approximated:

$$\frac{2\Gamma}{1+\Gamma \cdot a_{MSA}} \approx \frac{\kappa_D}{1+\kappa_D a_{DH}} \tag{30}$$

which is of course not the case when $a_{MSA} = a_{DH}$ and is approximate when $a_{DH}$ ($a$ applied in DH theory, which has the same value as $a_{MSA}$) is scaled by some factor $f_{sc}$ close to 5/6.[3] Scaling $a_{DH}$ by 5/6 in DH theory applies a reasonable correction for effects that arise from volumetric correlations between non-central ions, as exemplified for a non-central ion $j$ in the following figure:

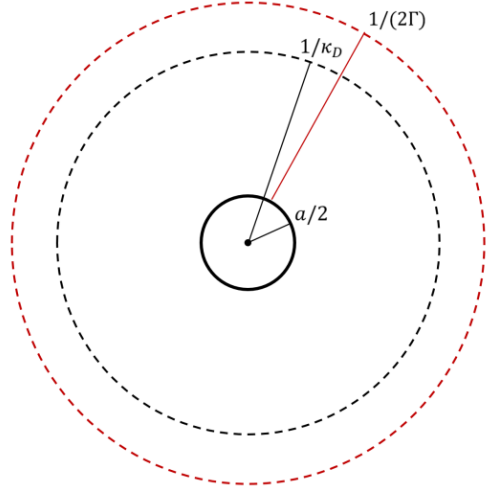

*Figure 5. A non-central ion j as handled by the DH and MSA theories.*

Based on Figure 5, one can guess from geometry that the difference between the theories equals a difference between the capacitance of the concentric spherical volumes given by the red (MSA) and black (DH) dashed lines. Therefore, from basic electrostatics the difference in capacitance between the two concentric spherical surfaces described by the inverse screening parameters is given by:

$$\Delta C^{MSA-DH} = \epsilon \left( \frac{1}{2\Gamma} - \frac{1}{\kappa_D} \right) \tag{31}$$

Equation (31) explicitly shows where the difference in capacitance comes from. One can arrive to the same conclusion based on a rearrangement of equation (30) and the concept of $f_{sc}$:

$$f_{sc} = \frac{a_{DH}}{a_{MSA}} = \frac{1}{2} + \left( \frac{1}{2\Gamma} - \frac{1}{\kappa_D} \right) \cdot \frac{1}{a_{MSA}} = \frac{1}{2} + \left( \frac{1}{2\Gamma} - \lambda_D \right) \cdot \frac{1}{a_{MSA}} \tag{32}$$

where we can see that $f_{sc}$ equals 1/2 of the salt diameter plus a correction of the difference in capacitance given by the screening lengths of both theories. Equation (32) has a limiting value of 1 at infinite dilution, an average value of 5/6 for low to moderate (5 M) concentrations and tends to values slightly lower than 5/6 at higher concentrations in aqueous systems.

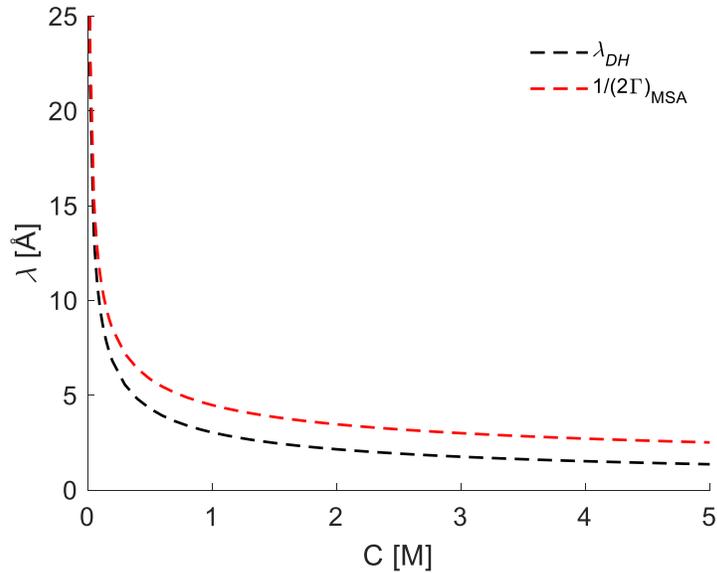

*Figure 6. Inverse screening parameters of DH and MSA theories. Calculations for $a = 4.25$ Å at 25°C with $\varepsilon_r = 78.65$.*

One question arises: why is an empirical scaling factor of $5/6$ such a good substitute for equation (32)? There are two central reasons for this.

Firstly, as the concentration rises, the effective capacitances given by the inverse screening parameters ($\lambda_D, 1/2\Gamma$) become a more or less linear function of concentration and so does the difference between them, as shown in Figure 6. On the other hand, at very low concentrations the ion size has no influence. Consequently, in the specific case of DH and MSA, the difference in effective capacitance given by $\frac{1}{2\Gamma} - \frac{1}{\kappa_D}$ is irrelevant at infinite dilution and becomes an almost linear function of concentration at higher ion densities, thus allowing for a simple constant scaling to represent this difference.

Secondly, the concentric spherical capacitors behave as in parallel in both theories[41,42] (see Figure 2). This allows for a direct additive contribution to the total effective capacitance when one of the capacitors is scaled: that is, an increase or decrease in the individual capacitance given by the screening parameter can be directly compensated by a decrease or increase in the individual capacitance of the salt diameter (central sphere). This finding is strictly limited to linearized theories.

The previous principle explains the workings behind the scaling of $5/6$ described by Maribo-Mogensen et al.[3] and can be applied to other linearized theories. For the PDH term, Pitzer[14,15,48,61] proposed a very reasonable average for practical application with $b = 2$, which is equivalent to $b = 14.9$ in the mole fraction scale assuming water as a solvent. Modifying the value of the closest approach parameter in the PDH term is the same as scaling the salt diameter and therefore also the same as changing the individual capacitance of the central sphere. This is due to the fact that the rest of the terms in $b$ reduce to universal constants and solvent properties as follows:

$$b = a \cdot \frac{\kappa_D}{\sqrt{I^{(m)}}} = a \cdot \sqrt{\frac{4\pi N_A \sum_i m_i q_i^2}{\epsilon k_B T}} \frac{1}{\sqrt{I^{(m)}}} = a \cdot \sqrt{\frac{2 N_A e^2 d_s}{\varepsilon_0 \varepsilon_r k_B T}} \tag{33}$$

where the density of the solvent $d_s$ comes from the common assumption in chemical engineering models that molarity equals molality: $c_i = m_i$. The PDH term with $b = 2$ yields very acceptable results for aqueous systems, but keeping $b$ as a constant for all systems possible is an inconsistent modelling strategy.

To describe the qualitative effect of overscreening with physically meaningful trends for the salt diameter $a$, a better semi-empirical closest approach parameter for the PDH term can be proposed. Based on our previous discussion this consists, once again, of a linear scaling of the salt diameter. As shown in Figure 7, the PDH term reproduces the qualitative behavior of the advanced MDE-DH theory fairly well by means of the following simple scaling for the physical closest approach parameter:

$$b = f_{sc} a \cdot \frac{\kappa_D}{\sqrt{I^{(m)}}} = 1.527 a \cdot \frac{\kappa_D}{\sqrt{I^{(m)}}} \tag{34}$$

This expression is exactly equivalent to scaling the salt diameter $a$ by a factor $f_{sc} \approx 3/2$ in equation (14). This scaling yields also reasonable qualitative results for 1:2 asymmetric and even for 1:3 asymmetric electrolytes. This is briefly shown in the Supporting Information.

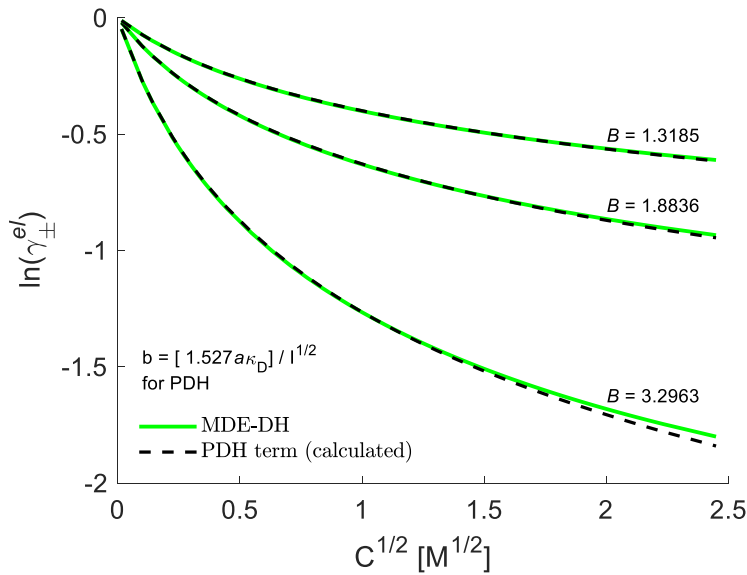

*Figure 7. Logarithmic ionic activity coefficients (electrostatic contribution only) calculated with the scaled PDH and with MDE-DH theory. Calculations for a salt diameter of 4.25 Å at three different fluid specifications (B). Calculations for the PDH term in the three cases assume $\rho_s = 1000 \, L/m^3$.*

By introducing the scaled salt diameter into the effective capacitance given by equation (14), the resulting function behaves very similarly to the effective capacitance of the MDE-DH theory from equation (10):

$$\left(\frac{1}{3}\right)\left[\left(\frac{\kappa_D}{1+\kappa_D f_{sc} a}\right) + \frac{2}{f_{sc} a} \ln(\kappa_D f_{sc} a + 1)\right] \approx \frac{1}{C^{MDE-DH}} \quad (35)$$

with $f_{sc} \approx 3/2$. The similarity of behavior holds below and above the Kirkwood cross-over for a very wide range of fluid specifications and for any value of $a$. Just like 5/6 in MSA and DH, these empirical scaling values are, to the best of our knowledge, not derivable in any fundamental context; yet their background becomes relatable through the point of view of capacitor circuit analogies.

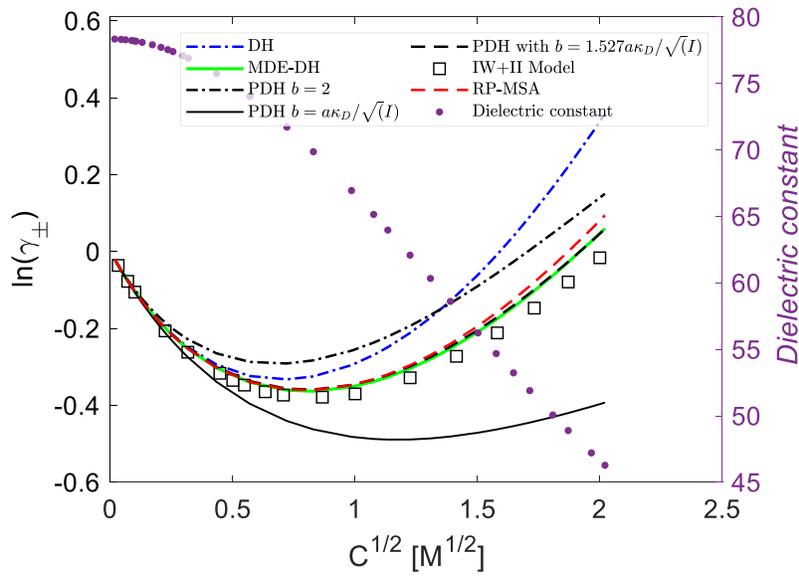

*Figure 8. Different models applied to aqueous NaCl ($a = 2.77$ Å) with a concentration dependent dielectric constant at 25°C. All linearized models for the electrostatic contribution are combined with the unrestricted hard sphere (UHS) contribution and a Born term. II+IW model values from Valiskó and Boda[50] are shown for mathematical reference.*

Figure 8 compares different electrolyte models for the specific case of aqueous NaCl and their comparison to a Monte-Carlo based method (IW + II Model[50]). Calculations with DH, MSA, PDH and MDE-DH are combined with a Born term and an unrestricted hard-sphere contribution according to equation (17). It can be observed for the PDH term that keeping the theoretical value of $b$ without scaling the diameter and that keeping $b$ as a constant are not good qualitative choices. With equation (34) it yields a virtually equal performance to MDE-DH when calculating the mean ionic activity coefficient. As expected,[43] MDE-DH lies closer to the reference values from the Monte-Carlo simulations when compared to DH.

If one averages the salt diameters of all alkali halides based on the values reported by Marcus[51] then one obtains a value that is close to 3.08 Å. This means that universal values between 4 Å and 4.5 Å for $a$ are qualitatively reasonable averages for equation (33) from a theoretical standpoint when performing calculations with the PDH term for simple salts and explains the partial success of alternative forms of $b$ with averaged values for $a$ in the literature[26,31,62] that reduce to $b \approx 6.4 \cdot \sqrt{d_s/(\varepsilon_r T)}$.

Moving on to considerations for underscreening, one may observe that as ion coupling increases due to low dielectric constant values, the recommended values[44] for $b$ increase dramatically. As a rule of thumb[63] for any 1:1 electrolyte, e.g. an average alkali halide ($a \sim 3.1$ Å), this can be significant when the dielectric constant is of $\varepsilon_r \leq 20$ at room temperature. This is for reference only, as in reality there is no hard boundary that separates the validity and non-validity of the LPB approximation based on $\varepsilon_r$ values. One extreme example is diethyl ether with $\varepsilon_r = 4.3$, where the LPB is not valid, yet thermodynamic quantities are commonly calculated adjusting $b \geq 90$, along with the other parameters of the Pitzer equations.[64]

These recommended values can be interpreted under the light of underscreening as an empirical adjustment of the salt diameter to compensate for a decreasing screening parameter. The experimental evidence[5,6,65–67] suggests the presence of considerable underscreening in highly concentrated electrolytes and ionic liquids, resulting in an anomalously large Debye length. This is associated with both a decreasing dielectric constant in the solution and, hypothetically, extensive ion pairing. Several theories have been invoked to explain this phenomenon,[5,49] for example the DIT and Bjerrum association theory. In the case of these two theories a central distinction is required: the Bjerrum based approaches theorize additional paired entities in the solution[68] (ion pairs, ion triplets,…) which may or may not be neutrally charged; whereas the DIT, an exact formalism, may contain these through many body correlations. A complete theory with sufficient pair correlations should treat all ions on the same basis, regardless of whether they are paired or not at a given instant.[10]

With these ideas, let us now propose a hypothetical capacitor circuit based on equation (2) that treats a central ion of radius $a/2$ and some hypothetical screening parameter $\kappa^H$ and assume this capacitor circuit can describe underscreening. We do not know yet how to calculate $\kappa^H$, but in analogy to equations (32) and (34), we can propose that the salt diameter in DH theory can be scaled by some function of $\kappa_D$ and $\kappa^H$ to acceptably approximate our hypothetical circuit configuration for underscreening. In light of the previous discussion for Figure 5 and Figure 7, we must approximate the difference between the effective capacitances. This approximation would then be given by:

$$\kappa_D a \cdot f(\kappa_D, \kappa^H) \approx \kappa_D a + \left(\frac{\kappa_D}{\kappa^H} - 1\right) \tag{36}$$

where $f(\kappa^H, \kappa_D)$, in direct analogy to equation (32), is a function that scales the salt diameter to make both theories behave as similarly as possible.

The major challenge now is to provide reasonable relations to approximate $f(\kappa_D, \kappa^H)$ in a physically meaningful manner. Several rules are self-evident: $f(\kappa_D, \kappa^H)$ must have a limiting value of 1 at infinite dilution and it must tend to relatively constant values as concentration rises. Forms of $f(\kappa_D, \kappa^H)$ therefore require approximations based on fundamental relations.

In the DIT, the variation of the leading screening parameter due to ion-ion correlations can be calculated at infinite dilution by the limiting law obtained from the total correlation function proposed by Kjellander and Mitchell:[46]

$$\left(\frac{\kappa}{\kappa_D}\right)^2 \sim 1 + \frac{\Lambda \ln(3)}{4}\left[\frac{\sum_i v_i z_i^3}{\sum_i v_i z_i^2}\right]^2 + \frac{\Lambda^2 \ln(\Lambda)}{6}\left[\frac{\sum_i v_i z_i^4}{\sum_i v_i z_i^2}\right]^2 + \mathcal{O}(\Lambda^2) \tag{37}$$

where $\Lambda = \kappa_D \lambda_B$ is a coupling parameter and the function $\mathcal{O}(\Lambda^2)$ can be approximated by $D\Lambda^2$ where $D$ is a constant.[43] The behavior described by this limiting law is supported by experimental evidence[6] and was approximated for symmetric electrolytes by the following lower order semi-empirical correlation that neglects the second and third terms in equation (37):[6]

$$\frac{\kappa}{\kappa_D} = \sqrt{1 + \alpha_1 \Lambda^2} \tag{38}$$

where $\alpha_1$ is a negative constant that correlates with the association constant and the Bjerrum length:[6]

$$\alpha_1 = -\frac{K_{assoc}(T, \varepsilon_r, a)}{8\pi z^2 N_A \lambda_B^3} \tag{39}$$

A major mathematical and theoretical hindrance for our practical purposes is that these relations are limiting laws. Furthermore, equation (38) becomes complex valued at higher concentrations. Nevertheless, a very satisfactory approximation for equation (38) at low concentrations is given by the following expression, which can also be semi-linearly extrapolated for higher concentrations:

$$\frac{\kappa_D}{\kappa} = \sqrt{1 - \alpha_1 \Lambda^2} \tag{40}$$

This approximation can be projected for higher concentrations without any mathematical obstacles and be treated semi-empirically. Our analogy automatically assumes that $\kappa^H \equiv \kappa$ is the only leading term. We assume that $\left(\frac{\kappa_D}{\kappa}\right)^2$ relates to a ratio of total ions to free (unpaired) ions as shown in the literature for the leading decay $\kappa$ in underscreening.[6,43]

With these considerations and assumptions, the introduction of equation (40) in equation (36) yields the following surprisingly simple form:

$$\kappa_D a \cdot f(\kappa_D, \kappa^H) \approx \kappa_D a + \left(\sqrt{1 - \alpha_1 \Lambda^2} - 1\right) \tag{41}$$

Equation (41) can then be directly introduced into the closest approach parameter in the PDH term as follows:

$$b = \frac{1}{\sqrt{I^{(m)}}} \cdot \kappa_D a \cdot f(\kappa_D, \kappa^H) \tag{42}$$

As stated before, $f(\kappa^H, \kappa_D) = \sqrt{1 - \alpha_1 \Lambda^2}$ attains a limiting value of 1, which makes equation (42) approach the conventional $b = \kappa_D a / \sqrt{I^{(m)}}$ for decreasing values of $\kappa_D$. Equation (42) by itself already provides a re-interpretation of the recommended[44] Pitzer values for $b$ based on principles from the DIT. Underscreening reduces the value of the screening length resulting in an anomalous decay length,[43,45] thus the hypothetical effective capacitance from $1/\kappa$ would become larger with underscreening. Because equation (42) is given in terms of $\kappa_D$ instead of $\kappa^H$, $f(\kappa_D, \kappa^H)$ must have an increasingly positive value to compensate. This is the verified trend in the recommended Pitzer values for $b$, which are considerably larger at low dielectric constant values. One could therefore argue that, historically, the recommended Pitzer values have been providing a scaling of the salt diameter to counter-balance the missing effects of an increase in effective capacitance that arises from underscreening.

To capture the qualitative effect of volume exclusion so that the PDH term performs similarly to MDE-DH when $f(\kappa_D, \kappa^H)$ is not dominant (e.g. in aqueous systems or at infinite dilution), we can use equation (34) to propose the following final semi-empirical rearrangement:

$$\kappa_D a \cdot f(\kappa_D, \kappa^H) \approx \frac{3}{2} a \kappa_D + \left(\sqrt{1 - \alpha_1 \Lambda^2} - 1\right) \tag{43}$$

The association constant $K_{assoc}$ may be estimated with some expression e.g. Fuoss,[69] Bjerrum[70] or Ebeling.[59] Furthermore, the salt diameter in these expressions has been historically treated as a fitting parameter that is usually larger than the sum of the lattice radii of the ions.[59,60,69–72] It is therefore convenient to use a universal fitting parameter $\alpha_{sc}$ in order to counterbalance the effect of excessively large values of $K_{assoc}$ when $\varepsilon_r$ is very low. The reproducibility of practical results can then be extended for application with g$^E$-models when some association constant (e.g. Ebeling[59]) is applied with $\alpha_{sc}$ as follows:

$$K_{assoc}(\alpha_{sc}, T, \varepsilon_r, a) = 8\pi N_A (\alpha_{sc} a)^3 \sum_{n=1}^{\infty} \frac{B_{sc}^{2n+2}}{(2n+2)! \cdot (2n-1)} \tag{44}$$

where the volume is given in m$^3$ ($a$ is given in 10$^{-10}$ m) and the fluid specification is scaled as follows:

$$B_{sc} = \frac{\lambda_B}{(\alpha_{sc}a)} = \frac{e^2}{\epsilon k_B T(\alpha_{sc}a)} = \frac{e^2}{4\pi\varepsilon_0\varepsilon_r k_B T(\alpha_{sc}a)} \tag{45}$$

Consequently, for a 1:1 electrolyte assuming molality equals molarity and introducing equation (43) into equation (42) with the previous definitions, one can reduce the result to the following explicit form for the closest approach parameter:

$$b = a\sqrt{\frac{9d_s N_A e^2}{2\varepsilon_0\varepsilon_r k_B T}} + \sqrt{\frac{1}{I^{(m)}} + d_s \cdot K_{assoc}(\alpha_{sc}, T, \varepsilon_r, a) - \frac{1}{\sqrt{I^{(m)}}}} \tag{46}$$

One may infer from equation (46) that $b$ approaches a constant value as ion concentration rises. The term $\frac{1}{I^{(m)}}$ and the term $\frac{1}{\sqrt{I^{(m)}}}$ could be neglected when a constant limiting value for $b$ is desired. This is however not done in the present work. The main reason for this is that the terms $\frac{1}{I^{(m)}}$ and $\frac{1}{\sqrt{I^{(m)}}}$ allow for a better description of the Debye slope at very low $\varepsilon_r$ values.

Using the radii from Marcus[51] we can estimate an average radius of $a = 3.08$ Å for alkali halides and with a value of $\alpha_{sc} \approx 4$, equation (46) reproduces trends that lie very near to the recommended literature values for the Pitzer equations, as shown in Table 1:

Table 1. Recommended values for $b$ from the literature as compared to the values calculated with equation (46) for a 1:1 electrolyte in several solvents with a wide range of dielectric constant values.

| Solvent ($\varepsilon_r$) | Recommended literature value[44,48] for $b$ | Equation (46) with $\alpha_{sc} = 4$ (0.001 M - 5 M with $a = 3.08$ Å at 25°C)* | |
|---|---|---|---|
| | | average value | [initial value – final value] |
| Water (78.34) | 1.2 | 1.5 | [1.5 – 1.6] |
| Acetonitrile (35.95) | 3.2 | 2.4 | [2.3 – 2.5] |
| Ethanol (24.35) | 3.2 | 3.2 | [2.8 – 3.4] |
| Dimethoxyethane (7.08) | 20 | 15.6 | [6.5 – 16.7] |
| Dimethyl Carbonate (3.11) | 95 | 90.8 | [54.6 – 92.4] |

*Concentration steps of 0.1 M above 0.1 M and concentration steps of 0.01 below 0.1 M.

Constant values of $\alpha_{sc} \approx 4$ and $a \approx 3.1$ Å are recommended as universal parameters for conventional (halides, sulfates, perchlorates, nitrates) symmetric and 1:2 asymmetric alkali salts. These can be applied to the modified closest approach parameter in the Pitzer equations or for effectively replacing the $b$ in the Pitzer-Debye-Hückel term of $g^E$-models like, for instance, AIOMFAC[26,62] or LIFAC/LIQAC.[27–29,31] In the upcoming section, this is implemented in the COSMO-RS-ES model[35–37] as a proof of principle.

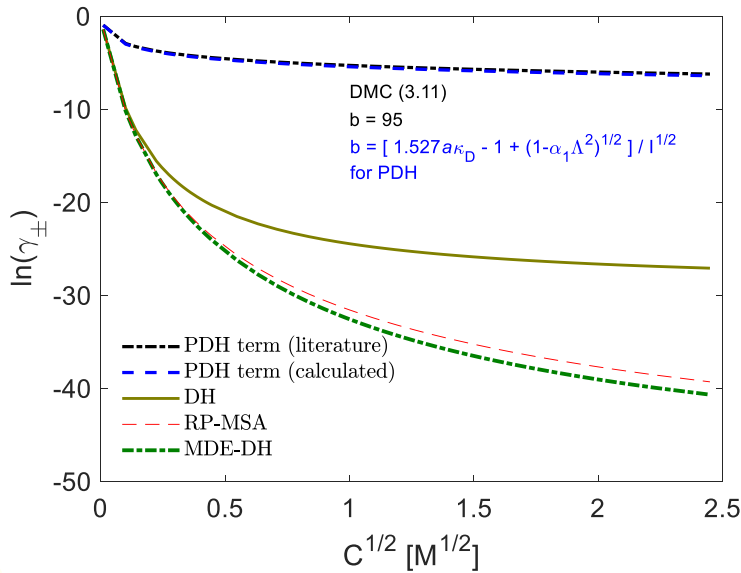

*Figure 9. Several electrostatic terms applied for $a = 3.08$ Å at 25°C in dimethyl carbonate (DMC) with $\varepsilon_r = 3.11$. Linear approximations are no longer valid and a consequent systematic underestimation of the activity coefficient is shown with MDE-DH, the restricted primitive MSA and Debye-Hückel. The PDH term is shown both with a recommended literature value[44] (black) and with equation (46).*

Figure 9 shows one case from Table 1. Similar complementary figures are included in the Supporting Information to show how at sufficiently low fluid specifications (low values of $\lambda_B/a$, i.e. aqueous electrolytes) equation (46) yields a behavior of the PDH term that is almost identical to MDE-DH, whereas at very high fluid specifications (high values of $\lambda_B/a$, i.e. ions in dimethoxyethane [DME] or dimethyl carbonate [DMC]) equation (46) reproduces recommended values for $b$. One can also observe the systematic underestimation for the linearized approaches (DH, MDE-DH and restricted primitive MSA) at very low dielectric constant values. This systematic underestimation caused large outliers in the calculations with previous versions of the COSMO-RS-ES model[35,36] (exemplified further ahead in Figure 10).

In the case of asymmetric 1:2 (or 2:1) electrolytes we take the same assumptions from Pitzer[14] as valid, but the effect of charge asymmetry is only retained in $\kappa_D$. No considerations for charge asymmetry have been introduced in $K_{assoc}$ or in $\lambda_B$. It is necessary to clarify that, while equation (38) has been used to approximate experimental data in 2:2 and 3:3 electrolytes,[6,43] equation (46) would not be applicable in these cases. The more extreme non-linear effects of highly charged systems and symmetric $n:n$ electrolytes with $n > 1$ cannot be captured properly by linearized electrolyte theories. It must be kept in mind that there may be more than one explanation for underscreening in an electrolyte[43] and we focus on addressing the specific case of underscreening in low dielectric constant mixtures for 1:1 electrolytes and our assumptions can acceptably stretch to 1:2 / 2:1 and perhaps even 1:3 / 3:1 electrolytes.

*Application with the COSMO-RS-ES model: results*

The COSMO-RS-ES model was originally developed with the PDH term using a constant value for the closest approach parameter.[35,36] This is not a good assumption when the dielectric constant attains low values and therefore in a previous publication[37] a correction was applied by means of an effective ionic strength from an iterative calculation based on the law of mass action and Bjerrum´s constant. It was then observed that this iterative calculation made the long-range term behave as if highly diluted when very strong ion-ion coupling in low dielectric constant solutions is expected and that this correlates with undersrceening.[37] These observations motivated replacing the computationally expensive iterative algorithm for ion pairing with more general semi-empirical expressions like the equation (46) to complement ion-ion interactions.

The short-range term from the COSMO-RS-ES equations already accounts for ion-ion contacts. Nevertheless, the long-range electrostatic term still needs an adjustment to counterbalance the systematic underestimation of linearized theories with very low dielectric constants.[37] This is addressed in the literature by the explicit inclusion of ion pairing[54,68,73,74] or partly addressed in the Pitzer equations by recommended values[44] for the closest approach parameter $b$. The former option has been tested[37] and with equation (46) we now attempt the latter strategy. For this purpose, COSMO-RS-ES was reparametrized replacing the constant $b = 14.9$ (the mole fraction equivalent of $b = 2$ in water) in favor of equation (46). The COSMO-RS-ES readjusted parameters after a single parameterization are shown in Table 2. The parameter $f_{COSMO}$, is also applied during the present work and remains in the vicinity of 0.82, keeping close similarity to the value of our previous publication.[37]

The only necessary change for implementation within COSMO-RS-ES is expressing equation (46) in the mole fraction convention:

$$b = a\sqrt{\frac{9d_s N_A e^2}{2M_s \varepsilon_0 \varepsilon_r k_B T}} + \sqrt{\frac{1}{I^{(x)}} + \frac{d_s}{M_s} \cdot 8\pi N_A (\alpha_{sc} a)^3 \sum_{n=1}^{\infty} \frac{B_{sc}^{2n+2}}{(2n+2)! \cdot (2n-1)}} - \frac{1}{\sqrt{I^{(x)}}} \quad (47)$$

where the salt diameter is given by the COSMO radii as $a = f_{COSMO}(r_+^{COSMO} + r_-^{COSMO})$ where $f_{COSMO}$ is a scaling parameter. This is only done for the first term. For the second term $\alpha_{sc} a = 12.4$ Å is intentionally fixed to support the argument that the literature values for $b$ indirectly approximate an underlying physical phenomenon as discussed in the previous section. The mole fraction convention also takes salt free properties for the molar mass of the system in addition to the $m_i = c_i$ assumption. This results in Gibbs-Duhem inconsistencies for molten salts and such a particular issue is not addressed in the present work. For orientation on this particular point, the reader is referred to the work of Chang and Lin.[39]

Table 2. Parameters for COSMO-RS-ES with the proposed modified PDH term.

| cation radii [Å] | | Parameters A,B $\left[\frac{kJ\,Å^2}{mol\,e^2}\right]$ | | Parameters C $\left[\frac{e}{Å^2}\right]$ | | parameters D,E [−] | |
|---|---|---|---|---|---|---|---|
| $Li^+$ | 1.712 | $A_1$ | 5277 | $C_1$ | 0.0062 | $D_1$ | 1852 |
| $Na^+$ | 1.880 | $A_2$ | 3876 | $C_2$ | 0.0099 | $E_1$ | 2.0747 |
| $K^+$ | 1.995 | $A_3$ | 3015 | | | $E_2$ | 7.33E-4 |
| $Rb^+$ | 2.073 | $A_4$ | 2564 | | | $E_3$ | 0.0698 |
| $Cs^+$ | 2.230 | $A_5$ | 3083 | | | | |
| | | $A_6$ | 3990 | | | | |
| | | $B_1$ | 13866 | | | | |
| $f_{COSMO}$ | 0.825 | $B_2$ | 337 | | | | |
| $\alpha_{sc}a$ taken as constant | 12.4 [Å] | $B_3$ | 5037 | | | | |
| | | $B_4$ | 35 | | | | |
| | | $B_5$ | 22401 | | | | |
| | | $B_6$ | 1004 | | | | |
| | | $B_7$ | 12060 | | | | |
| | | $B_8$ | 16986 | | | | |

Table 3. Comparison of different results (AAD: Average Absolute Deviation) with the COSMO-RS-ES model with a two-step parameterization.

| | Data Set ($N_{DP}$) | COSMO-RS-ES (original)[36] | COSMO-RS-ES (+ ion pairing)[37] | COSMO-RS-ES (+ modified PDH)[this work] |
|---|---|---|---|---|
| | $b$ | 2* | 2* | equation (46) |
| Part of the training set | MIAC aq. (1206) | 0.043 | 0.043 | 0.040 |
| | LLE (1056) | 0.675 | 0.518 | 0.531 |
| | GT (992) | 3.158 | 3.768 | 3.865 |
| Predictions | SLE (836) | 1.000 | 0.896 | 0.954 |
| | MIAC org. (980) | 2.679 | 1.274 | 1.008 |

*The mole fraction equivalent for aqueous systems of $b^{(m)} = 2$ is $b^{(x)} = 14.9$, commonly taken as a universal value. MIAC aq. = aqueous MIAC values; LLE = liquid-liquid equilibria; GT = Gibbs free energies of transfer of single ions from water to mixed and pure organic solvents; SLE = solid-liquid equilibria; MIAC org. = MIAC values in pure organic solvents.

Different results are obtained for the COSMO-RS-ES versions presented in Table 3. The two modified versions outperform the original version[35] whereby both apply a correction to the long-range term for situations where the DH theory is not valid: one with an effective ionic strength and the other with a modified PDH term. The present work provides the best description for aqueous and non-aqueous MIACs, while COSMO-RS-ES + ion pairing still allows marginally better quantitative SLE predictions and LLE correlations. In general terms, both version of the model perform very well.

The predictions of the SLE and LLE systems in this work are almost the same as with the COSMO-RS-ES version + ion pairing with minor qualitative differences. However, such qualitative differences are better appreciated when evaluating the MIAC predictions of different salts in pure organic solvents, which are perhaps some of the most challenging systems for any predictive electrolyte model.

As can be seen in Table 3 the modified PDH term improves the description of this type of data by about 20%. Two examples are shown in Figure 10 for these challenging systems were strong ion coupling can be expected: LiClO$_4$ in pure dimethoxyethane ($\varepsilon_r = 7.08$) and NaI in 2-propanol ($\varepsilon_r = 20.18$). As previously shown in Figure 9 for DMC, linearized electrolyte theories exhibit a considerable underestimation in the order of several logarithmic units for low $\varepsilon_r$ values. The deviation would be even larger with the PDH term if $b$ is left with a constant value of 2 (see Figure 10 - left). This underestimation is roughly corrected with an effective ionic strength by our previous COSMO-RS-ES + ion pairing approach. In contrast, the qualitative behavior is much better described by the modified PDH term, as shown in Figure 10 for the aforementioned systems.

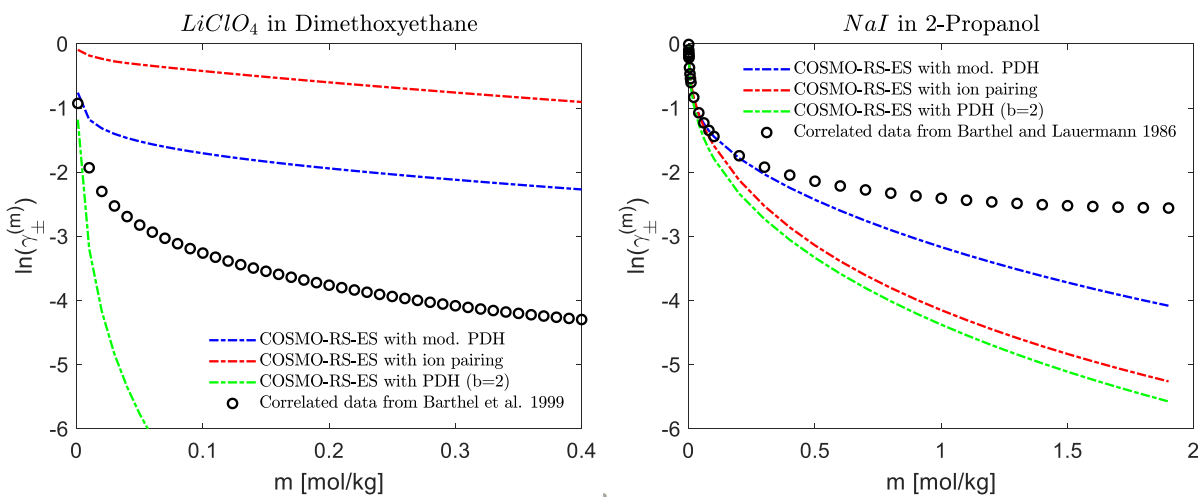

*Figure 10. Predicted logarithmic activity coefficients based on different COSMO-RS-ES parameterizations (unmodified PDH term [green][36], PDH term with ion pairing [red][37] and modified PDH term [blue], this work). The black dots were calculated from the correlated Pitzer models based on the experimental data from Barthel et al.[64] for the system LiClO$_4$ in pure dimethoxyethane ($\varepsilon_r = 7.08$) and from Barthel and Lauermann[75] for the system NaI in pure 2-propanol ($\varepsilon_r = 20.18$).*

The previous observations show that the model has been made more consistent for low to moderate concentrations in aqueous, mixed organic and pure organic systems while also retaining qualitative corrections for strong ion coupling in 1:1 and 1:2 electrolytes. Furthermore, the model no longer needs an iterative calculation of the effective ionic strength but approximates the qualitative effect of underscreening with the modified PDH term.

Systematic deviations as salt concentration rises considerably in organic solvents are expected. These deviations are traditionally addressed either by the introduction of physical phenomena like the description of the dielectric constant of the solution[18,22,39] or by the introduction of additional semi-empirical parameters.[31,76,77] In this regard, the objective of the current work was to replace an iterative ion pairing algorithm with simpler, universally applicable and more consistent semi-empirical relations that can be tracked to a physically significant origin. Subsequent revisions and extensions of the COSMO-RS-ES short-range equations will be the objective of future work.

## Conclusions

The present work retakes the capacitor analogy for the restricted primitive model from Lesser Blum[42] in order to present its general applicability in various electrolyte theories. Through capacitor circuit representations of different theories an explanation is offered of why scaling the salt diameter blurs differences between them when calculating practical thermodynamic quantities. The central concept behind this lies in the parallel configuration of the individual capacitances when the theories are represented as circuits. The ion cloud and the central sphere behave as capacitors in parallel and, consequently, their additive nature makes it possible to scale the diameter as a good empirical approximation for the missing volume effects and over-screening. This principle is then extrapolated through an analogy applying an approximation for the Dressed Ion Theory to modify the Pitzer-Debye-Hückel term. By these means we obtain a surprisingly simple function for the closest approach parameter that is in qualitative and quantitative agreement with recommended literature values for the Pitzer equations.[44]

This transferable modification of the PDH term (applicable to any $g^E$-model for electrolytes) is presented in order to facilitate the understanding and potentially the better modelling of thermodynamic activity and phase equilibria of electrolyte systems in chemical engineering. We expect that a more balanced long-range term with a qualitative correction that allows a wider range of applicability will facilitate that the short-range terms account for ion-solvent and ion-ion interactions in a more flexible, efficient manner.

In our previous work[37] a correlation between the effective ionic strength of a solution with paired ions and the physical principles of underscreening had been suggested. The modified PDH term presented here is the end result of that particular observation. It is successfully applied to the predictive COSMO-RS-ES model, an own reimplementation of the COSMO-RS theory especially improved for thermodynamic property calculation of electrolyte systems. The new term is an effective replacement of the more elaborate and computationally expensive ion pairing approach, as well as a choice that provides a better description of the Debye slope in a wider scope of solvents (when classified by the dielectric constant). This exercise provides evidence that the recommended literature values for the closest approach parameter have an underlying physical meaning and can now be predicted with simple relations.

# Acknowledgements

AGC would like to express gratitude to Roland Kjellander for the enlightening discussions on electrolyte theory and for the very constructive comments on the manuscript of this paper. The authors also thank Thomas Gerlach for a brief discussion that reinforced some of the ideas presented here and Dezsö Boda for providing the data necessary for Figure 8.

# Conflicts of Interest

There are no conflicts of interest to declare.